# Ultrafast Dynamics Evidence of High Temperature Superconductivity in Single Unit Cell FeSe on SrTiO$_3$

Y. C. Tian[1], W. H. Zhang[2], F. S. Li[2], Y. L. Wu[1], Q. Wu[1], F. Sun[1], Lili Wang[2,3], Xucun Ma[2,3,*], Qi-Kun Xue[2,3,*], Jimin Zhao[1,*]

[1]*Beijing National Laboratory for Condensed Matter Physics and Institute of Physics, Chinese Academy of Sciences, Beijing 100190, China*

[2]*State Key Laboratory for Low Dimensional Quantum Physics and Department of Physics, Tsinghua University, Beijing 100084, China*

[3]*Collaborative Innovation Center of Quantum Matter, Beijing 100084, China*

* Corresponding author.

Email: xucunma@mail.tsinghua.edu.cn; qkxue@mail.tsinghua.edu.cn; jmzhao@iphy.ac.cn

**The recent observation of high-temperature superconductivity in one unit cell thick FeSe on SrTiO$_3$ (1UC FeSe/STO)[1] initiated a new pathway to realizing even higher transition temperature ($T_c$) superconductors and established a promising different paradigm to understand the mechanism of high temperature superconductivity[2-10]. However, independent confirmation on the high $T_c$ value, the role of capping layer in reducing $T_c$, and the underlying mechanism of the novel superconductivity remain open questions[11]. Here we report a time-resolved study of the excited state ultrafast dynamics of 1UC FeSe/STO protected by 2UC FeTe capping layer, and identify the superconducting $T_c$ to be 68 (-5/+2) K. We found a coherent acoustic phonon mode in the capping layer, which provides an additional decay channel to the gluing bosons and explains the reduced $T_c$ value. Our investigation supports the phonon pairing scenario.**

The 1UC FeSe/STO[1] with high $T_c$ has recently attracted much attention. The $T_c$ values have been reported to be 55±5 K[2], 65±5 K[3], 60±5 K[4], and 55 (-5/+9) K[5] in angle-resolved photoemission spectroscopy experiments, 40 K[1,6], 109 K[7], and 80 K[8] in transport measurements including the Meissner effect, respectively. Electron-phonon coupling has been proposed[1] and analyzed to play a key factor for the observed high $T_c$ value[5,9,10]. While the in-consistency in the $T_c$ values needs to be clarified, more importantly, the superconducting nature of the phase transition needs to be understood comprehensively by using different experimental methods[11].

As a powerful means to investigate ultrafast dynamics of excited states, gap-perceptive ultrafast spectroscopy provides information on comprehensive interactions among the quasiparticles and various excitations, such as electron-phonon (*e-ph*) coupling, and has been used to probe and control novel quantum materials and superconductors[12-17]. In this work, we apply ultrafast spectroscopy (see Methods) to 1UC FeSe/STO system, where we demonstrate that it is single-layer sensitive. The superconducting (SC) $T_c$ value, *e-ph* coupling strength $\lambda$, SC energy gap $\Delta(T)$, density of thermal electrons $n_{th}$, as well as electron temperature $T_e$ have been obtained. Furthermore, we coherently generate and detect an acoustic phonon mode in the thin capping layer of FeTe, which clearly reveals a side-decay channel to the gluing bosons, hence explains the reduction of $T_c$ due to the protecting layer. Our results support the scenario of electron-phonon coupling.

Figure 1 summarizes the normalized photo-induced transient differential reflectivity, recorded from the sample structure schematically illustrated in the inset

(the detail of the growth procedure is described in Refs. 1,6). It can be seen that the scanning traces are composed of both increasing and decreasing components, whose amplitude and lifetime vary with temperature. Prominently, a hump is seen in the scan traces after the time-zero. The magnitude of the hump decreases and then increases as temperature rises up, showing a change around 70 K (inset of Fig. 1). This is related to the SC transition as discussed below. Our sample consists of only a single UC thick FeSe film covered by 2UC thick capping layer of FeTe to prevent air contamination. In our control experiment (Supplementary Information S1) we found that 2UC FeTe layer is of optimized thickness.

To identify phase transitions through ultrafast carrier dynamics, we map out the transient reflectivity as a function of both time and temperature in Fig. 2a. A few temperature regions can be seen, corresponding to different phases. Particularly, both the fast (at 0~2 ps, in white and purple colors) and slow (at > 2 ps, in red and blue colors) components show distinct changes around 68 K (Fig. 2a). In ultrafast spectroscopy this corresponds to a change in carrier density and lifetime, intrinsically correlated to a phase transition. In Fig. 2b we show the integral of ΔR/R within the initial 1.4 ps (*i.e.* the fast component in Fig. 2a) as a function of temperature. A clear decrease of the integral can be seen, which ends at 68 K (red dashed line in Fig. 2b). For ultrafast spectroscopy of cuprates, this corresponds to the SC $T_c$. We further show a different (quantitative) data analysis below, which unambiguously demonstrates that this is a SC transition, with a high $T_c$ of 68 (+2/-5) K.

The time-resolved processes can be approximated by two exponential components: $\Delta R/R = A_{fast}\exp(-t/\tau_{fast}) + A_{slow}\exp(-t/\tau_{slow}) + A_0$, where $A_{fast}$, $\tau_{fast}$, $A_{slow}$, and $\tau_{slow}$ are functions of temperature, which are obtained by fitting the experimental data (Fig. 3). It can be seen that $A_{slow}$ abruptly decreases around 68 K (Fig. 3c) as temperature increases. Moreover a $\lambda$-shaped distinct change in $\tau_{slow}$ is clearly seen at a slightly lower temperature than 68 K. The simultaneous observation of the two prominent changes strongly suggests a SC phase transition[13,15,16,19]. Note that usually $T_c$ is slightly higher than the temperature corresponding to the $\lambda$-shaped lifetime singularity. Microscopically, the photo-excited quasi-particles far above the Fermi surface relax to right above the SC gap and form a quasi-equilibrium quantum mixture with the high frequency phonons (with $E_{ph} > 2\Delta$, where $\Delta$ is the SC gap). The balance between the formation and deformation of Cooper pairs evolve in a way dictated by the phonon population—the so-called phonon bottleneck effect. This effect is most directly observed using time-resolved ultrafast spectroscopy and is well described by the Rothwarf-Taylor model[18]. The effect we observe is reflected in Fig. 2a (long vertical blue stripe nearby the white line) and Fig. 3d (the large $\tau_{slow}$ value around 60 K).

Quantitatively, we theoretically simulate the results in Fig. 3 by the model used by Kabanov *et al.* for a Cooper-pairing superconductor[13]. The gap-sensitive nature of the ultrafast dynamics is reflected, for an *e-boson* interacting system, through the temperature-dependent differential reflectivity as

$$|\Delta R/R| \propto \frac{\varepsilon_I/[\Delta(T) + k_B T/2]}{1 + \frac{2\nu}{N(0)\hbar\Omega_c}\sqrt{\frac{2k_B T}{\pi\Delta(T)}}e^{-\Delta(T)/k_B T}}, \qquad (1)$$

where $\varepsilon_I$ is the absorbed laser energy density per unit cell, $\Omega_c$ is the cut-off frequency of the phonons, $v$ is the effective number of interacting phonon modes per unit cell, $N(0)$ is the density of states at the Fermi surface, and we have assumed symmetric gap $\Delta(T) = \Delta(0)[1-(T/T_c)]^{1/2}$. With Eq. (1) we fit the data in Fig. 3c and find that the fitting curve for $A_{\text{slow}}$ has a sharp decrease ending at 68 K. As for other conventional, cuprate, and iron-based superconductors, this ending temperature is the $T_c$ of a SC phase transition[13,16,19]. In parallel, using the same model, the gap-dependent lifetime of the quasiparticles (and phonons in equilibrium) can be described as[13,19]

$$\tau_{slow} = \tau_{ph} = \frac{\hbar\omega^2 \ln\{1/[\varepsilon_I / 2N(0)\Delta(0)^2 + e^{-\Delta(T)/k_B T}]\}}{12\Gamma_\omega \Delta(T)^2}, \quad (2)$$

where $\omega$ is the high frequency phonon frequency and $\Gamma_\omega$ is the phonon linewidth. With Eq. (2) we fit the $\tau_{slow}$ data in Fig. 3d, which exhibits a $\lambda$-shaped singularity around 60 K. Multiple tries of the fitting parameters give a 63 K < $T_c$ < 70 K bound, with the best optimized value of $T_c$ = 68 K.

Simultaneously the above analysis yields $\Delta(0) = 3.2 \sim 3.7\, k_B T_c$ (i.e. 20.2±1.5 meV, taking the medium value), which is in excellent agreement with the reported 20.1 meV[1] and 3.0~3.5 $k_B T_c$[3] values. Also obtained is the density of thermal quasiparticles $n_{\text{th}}$ (inset of Fig. 3c), which increases prominently as temperature approaches $T_c$. By fitting $\tau_{slow}$ we obtain $\varepsilon_I / 2N(0)k_B^2 = 1.93 \times 10^4 (\text{K}^2)$, from which we derive that $N(0)$ is 69 state eV$^{-1}$spin$^{-1}$cell$^{-1}$, with $\varepsilon_I$ being $0.63 \times 10^{-20}$ Jcell$^{-1}$. This is more than 10 times larger than the values for typical cuprates and bulk iron-based superconductors. Moreover, by fitting $A_{slow}$ we obtain $2v/[N(0)\hbar\Omega_c] = 11.89$, and by using the cutting-off phonon energy $\hbar\Omega_c = 106$ meV for STO[20] we calculate that $v = 43$. This is also much higher than that for bulk FeSe[21], 1.6~2. Therefore, it is understandable that higher $N(0)$ and $v$ values lead to enhanced *e-ph* interaction (*i.e.* more electrons and phonons condense to the SC ground state), thus enhancing the $T_c$. Note that due to the oxygen vacancy induced (2×1) reconstruction at the FeSe-STO interface[22], the interface unit cell assumes a complex form of $Se_2Sr_4Ti_6O_{10}$, which allows for 63 total optical phonon modes. This is more than the 12 modes for bulk perovskite STO and makes $v = 43$ plausible.

We can directly obtain $\lambda$ from the quasiparticle lifetime. The relaxation dynamics is determined by the *e-ph* coupling with $\lambda\langle\omega^2\rangle = \frac{\pi}{3}\frac{k_B T_e}{\hbar\tau_{e-ph}}$, where $\tau_{e-ph}$ can be directly measured by our pump-probe spectroscopy ($\tau_{fast} = 0.23$ ps, see the red dots in Fig. 3b, which is the de-convolution value of raw $\tau_{fast}$) and $\lambda\langle\omega^2\rangle$ is the second moment of the Eliashberg function[23]. From the laser fluence we estimate that $T_e = 902$ K (Supplementary Information, S2). We speculate that the *e-ph* coupling is mainly between the FeSe electron and the interface and bulk STO phonons, owing to the vast atoms contributing to the phonon bath. Taking the experimentally measured low optical phonon frequency near Γ for bulk STO[20], $\hbar^2\langle\omega^2\rangle = 22^2 (\text{meV})^2$, we obtain $\lambda = 0.48$, which is in excellent agreement with the value of 0.5 in angle-resolved photoemission spectroscopy measurement[5]. Comparing with the $\lambda$ (0.16) for bulk FeSe[15], *e-ph* coupling plays a much more crucial role in the SC mechanism for 1UC FeSe.

To address whether phonon is the pairing glue and why a capping layer leads to the reduction of $T_c$[11], we investigate the effect of the capping layer. We prepared a separate sample of 10UC FeTe/STO, where we unambiguously generated and detected a coherent acoustic phonon mode[17,24,25] in the FeTe layer (Fig. 4a). A clear periodic phonon oscillation superimposes on a single-exponential electronic decay. In the lower panel coherent acoustic phonon waves are explicitly displayed, and in the inset the phonon frequency of 0.052 THz (0.22 meV) is obtained by Fourier transformation. Identical acoustic phonon mode is also observed in a 2UC FeTe/STO sample (Supplementary Information, S3). Observing such coherent acoustic phonon mode is non-trivial, considering that there are lots of islands and patches for the capping FeTe layer.

We contemplate that this acoustic phonon mode in the FeTe capping layer forms a decay channel for the gluing bosons (*e.g.* the optical phonons at the interface and in bulk STO) in the SC condensate, and thus reduces the $T_c$ of the 1UC FeSe/STO sample[11]. This scenario is schematically illustrated in Figs. 4b & 4c. In Fig. 4b optical phonon in STO and interface tunnels through the single FeSe layer and decays into the acoustic phonons in the FeTe capping layer. Although other origins such as defects, impurities, interference, and magnetic scattering may also play a role in reducing $T_c$, anharmonic decay of the optical phonons into acoustic phonons is both intrinsic and efficient[26], thus likely plays a more crucial role. In Fig. 4c we schematically show in the momentum space the additional decay channel due to the FeTe acoustic phonon. With such a scenario, corroborated by strong *e-ph* coupling addressed above, our observation of the acoustic phonon mode in the FeTe layer provides a reliable experimental support to the phonon-pairing mechanism. For example, it is consistent with the recent observation of the phonon replica[5]. Significantly, the anharmonic decay of the optical phonon into more channels of acoustic phonon modes will inevitably result in additional broadening of the phonon mode[26], thus an associated less sharp SC transition. Comparing the two transport measurements in Ref. 1 and 7, this is exactly what has been observed.

Moreover, we note that in Fig. 4c the SC gap has a value that is very close to the bottom of the optical phonon dispersion. This lead us to speculate that: The minimum value of the optical phonon may be the underlining limit for the SC gap, with $\Delta(0) \leq \text{Min}\{\hbar\Omega_{optical\ phonon}\}$, and thus an upper limit for $T_c$ (through $\Delta(0) = ck_BT_c$, with a coefficient $c$ contingent on the nature of superconductivity). Suppose that $\Delta(0) > \text{Min}\{\hbar\Omega_{optical\ phonon}\}$, optical phonons may lose energy to generate electronic excitations with intra-gap energies, which makes the electronic continuum at $\Delta(0)$ non-equilibrium. At the energy crossover this leads to the possible emergence of Higgs boson collective mode[27] (not necessarily be restricted to charge density wave) between optical phonons and the SC order parameter (gap). The ultimate result is lowering $\Delta(0)$ to a level below all the optical phonon energies. Although our investigation does not exclude other gluing mechanism, optical phonons can still be a limit for the $\Delta(0)$ and $T_c$ in those scenarios. This is because many of the suggested gluing excitations could decay through interacting with optical phonons. As for lowering $T_c$ values for other suggested gluing bosons, the same is true for the role of the acoustic phonon in the capping layer.

Besides, we have also performed power-dependence investigation (Supplementary Information, S4), which indicates that the 1UC FeSe superconductor is nodeless[14], being consistent with the observation in angle-resolved photoemission spectroscopy[28]. We ascribe the 103 K peak in Fig. 3d to the antiferromagnetic to paramagnetic phase transition of the FeTe layer (Supplementary Information, S5). And this demonstrates that our method is single-layer sensitive and might also apply to other interface investigations[29].

Significantly, the $T_c$ value we have obtained from the current ultrafast dynamics investigation is the highest for 1UC FeSe samples with protecting layer. It is higher than those $T_c$ values obtained using other methods[1-6]. This is because our method is non-invasive to the sample and single-layer sensitive. Without the capping layer, we expect that our ultrafast spectroscopy will give higher values of $T_c$ for an *in situ* measurement.

In concluding remark high temperature SC transition with $T_c$ = 68 (-5/+2) K has been observed in a 1UC FeSe/STO sample using ultrafast spectroscopy. This is the highest $T_c$ value observed in similar samples with capping layer. Gap energy of $\Delta(0)$=3.2~3.7 $k_B T_c$ is observed and strong e-ph coupling is identified with $\lambda$=0.48. Our observation of a coherent acoustic phonon mode in the FeTe capping layer explains why a capping layer reduces the $T_c$ value and suggests a phonon-pairing mechanism. Furthermore, an optical phonon limit to SC gap and $T_c$ is proposed. Our experimental results support that phonon plays an essential role in this new high $T_c$ SC system and our investigation paves way for non-contact investigation of single-layer or interface quantum materials.

## Methods

**Time-resolved ultrafast spectroscopy experiment.** Typical ultrafast spectroscopy experiment was carried out in the reflection geometry and with balanced detection. Ultrafast laser pulses of 800 nm central wavelength, 96 fs pulse width, and 250 kHz repetition rate were used to investigate the ultrafast dynamics of the 1UC FeSe on insulating STO sample from 5 K to room temperature. The relatively low repetition rate is used to avoid steady state laser heating, while maintaining significant signal-to-noise ratio. We measured the differential reflectivity, $\Delta R/R$, which is proportional to the excited state quasiparticle density in the sample. To avoid thermal effect due to laser heating, the pump and probe beam fluences were kept at 268 $\mu$J/cm$^2$ and 30 $\mu$J/cm$^2$, respectively. Cross-polarization detection was implemented to further enhance the signal-to-noise ratio.

## Acknowledgements

This work was supported by the National Basic Research Program of China grant (No. 2012CB821402, No. 2015CB921001), the National Natural Science Foundation of China grants (No.11274372, No.11374336) and the External Cooperation Program of the Chinese Academy of Sciences grant (No. GJHZ1403).

## Author contributions

J. M. Zhao and Q. K. Xue proposed and designed the research. X. C. Ma, L. L. Wang, W. H. Zhang, and F. S. Li contributed to MBE thin-film preparation. Y. C. Tian, Y. L. Wu, Q. Wu, F. Sun, and J. M. Zhao contributed to the ultrafast spectroscopy measurement and analyzed the data. J. M. Zhao, X. C. Ma, and Q. K. Xue wrote the paper.

# Figures and captions

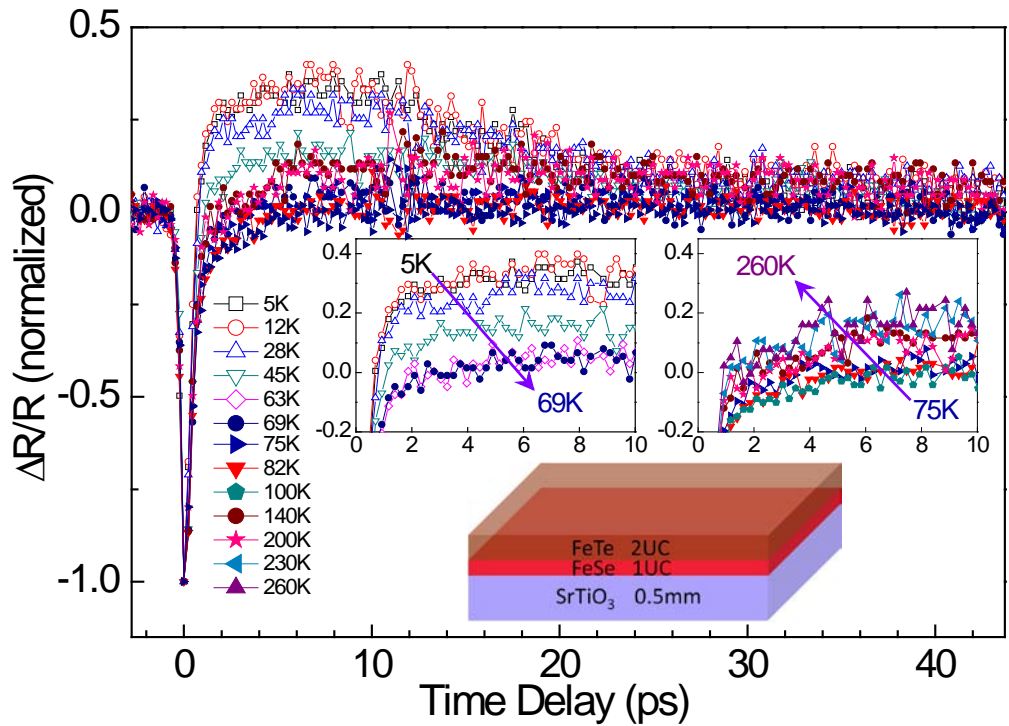

**Figure 1 | Normalized photo-induced differential reflectivity**. Scanning traces at different temperatures are illustrated, with zoom-in view shown in insets. The purple arrows mark the hump varying trends on temperature increase, respectively. Also shown is the schematic sample composition, where UC represents unit cell.

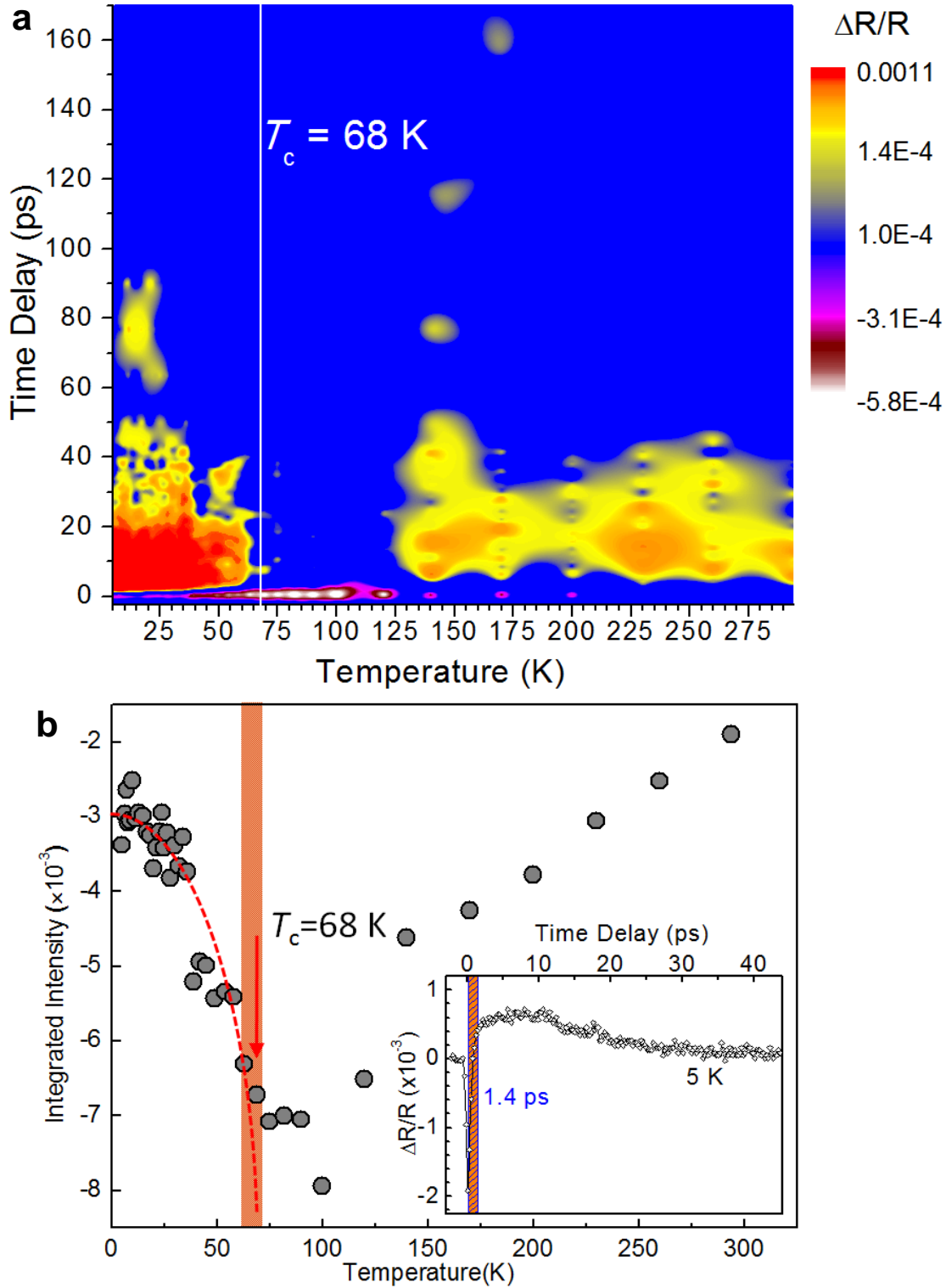

**Figure 2 | Qualitative evidences for SC transition at 68(-5/+2) K. a**, Two dimensional color mapping of the differential reflectivity as a function of both temperature and time. The white line marks the SC $T_c$. **b**, Temperature dependence of the integral of ΔR/R values within the initial 1.4 ps. The orange color stripe marks the $T_c$ range of 63~70 K. The inset shows the integration region.

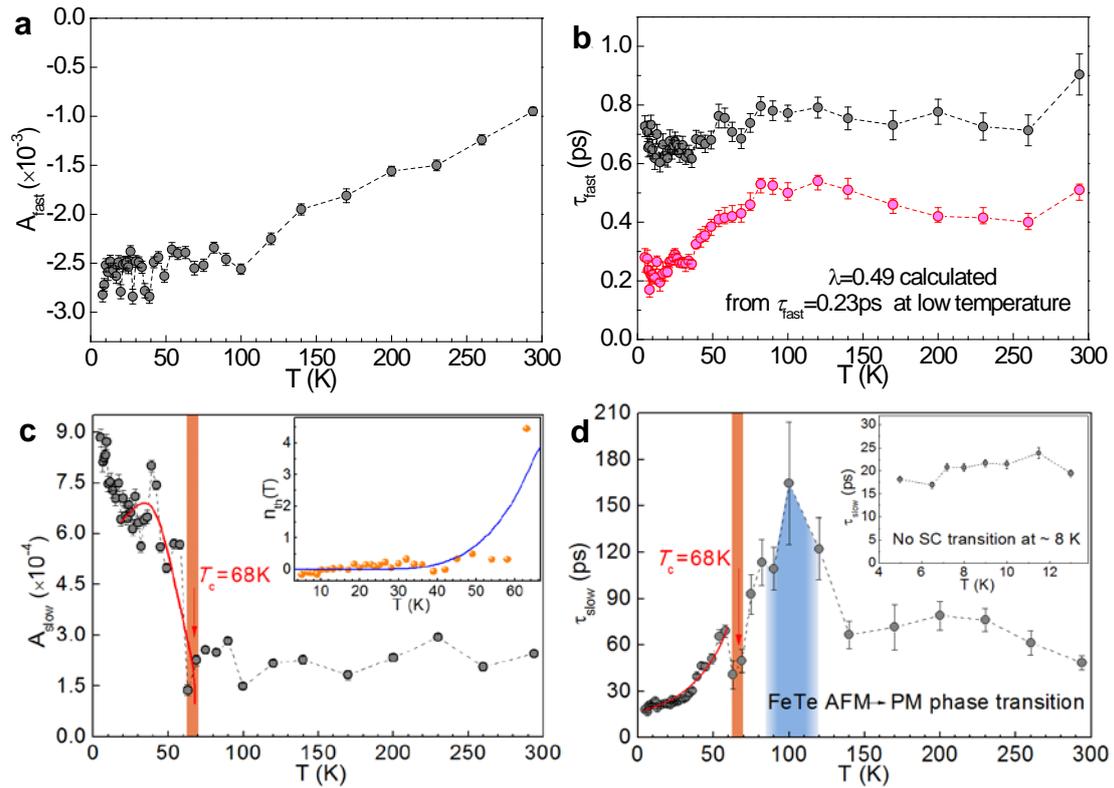

**Figure 3 | Quantitative evidences for SC transition at 68(-5/+2) K. a and b,** Temperature dependence of the amplitude and lifetime of the fast component. The red dots in **b** are the de-convoluted data. **c and d**, Temperature dependence of the amplitude and lifetime of the slow component. The red lines in **c** and **d** are simulation curves based on the Cooper-pairing theoretical model. The orange stripes mark the $T_c$ value range of 63~70 K. The light-blue stripe marks the phase transition in FeTe capping layer. Inset: Density of thermally-excited QPs (**c**) and Zoom-in view of the data around 8K (**d**).

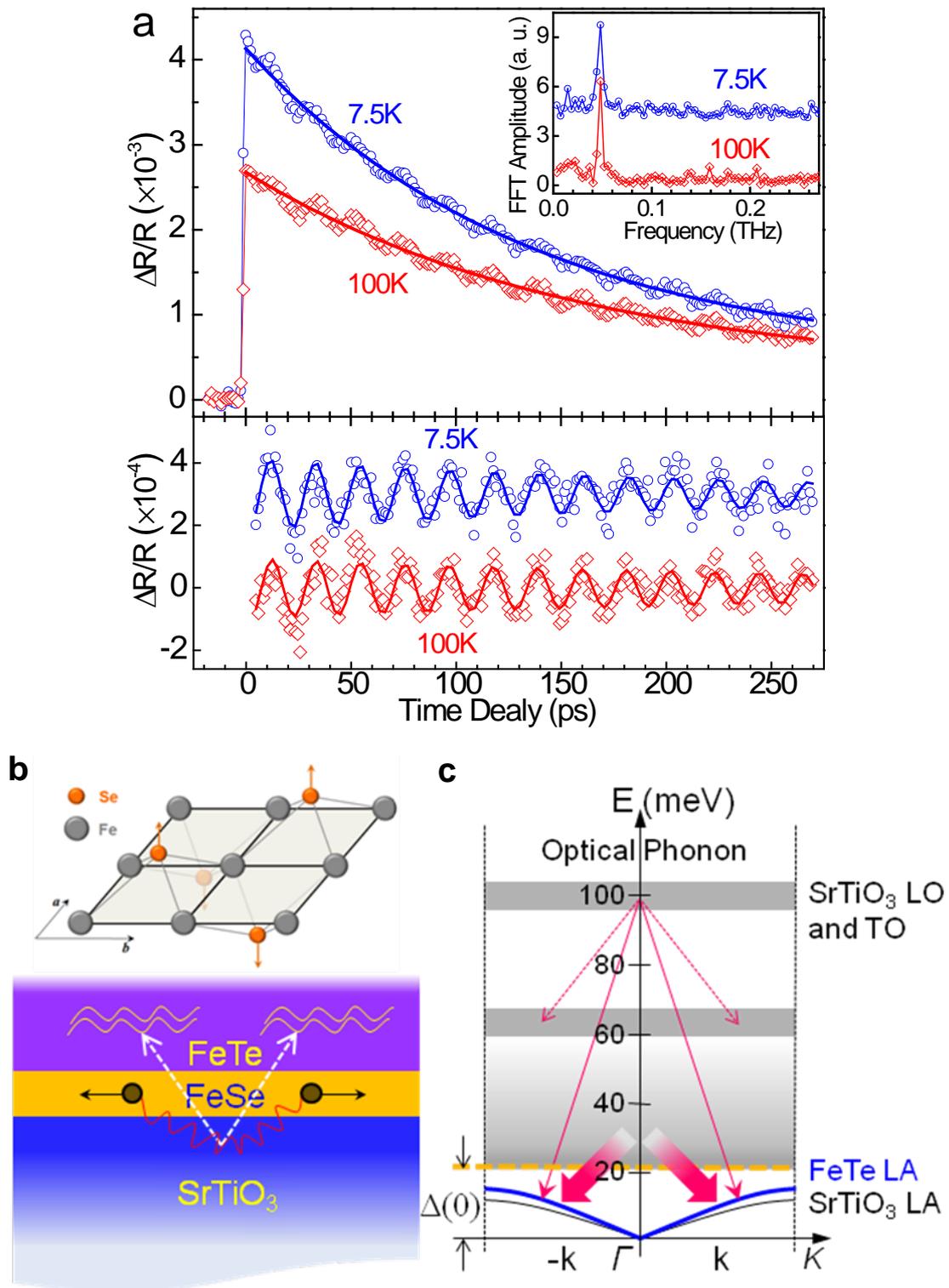

**Figure 4 | Essential role of phonon in the SC transition. a**, Ultrafast dynamics of the 10UC FeTe on SrTiO$_3$ at 7.5 K (blue) and 100 K (red), respectively. The thicker solid curves are single exponential decay fits. The oscillation components are the coherent acoustic phonon mode, which is displayed in the lower panel. Inset: Fourier transformation showing the phonon frequency. **b**, Schematic of the phonon decay tunneling through the FeSe layer.

The red wave represents optical phonon in the STO substrate and at the interface, while the yellow waves illustrate the acoustic phonons in the FeTe capping layer. Also shown is the schematic 1UC FeSe atomic structure. **c**, Schematic K-space diagram showing the decay of STO optical phonon into the additional FeTe acoustic phonon (blue dispersion curve). The red arrows represent the additional new channels of decay. The SC gap Δ (0) is also marked to scale. The gray bars represent roughly the optical phonon modes, with deeper color for higher density.

Supplementary Information

# Ultrafast Dynamics Evidence of High Temperature Superconductivity in Single Unit Cell FeSe on SrTiO$_3$

Y. C. Tian[1], W. H. Zhang[2], F. S. Li[2], Y. L. Wu[1], Q. Wu[1], F. Sun[1], Lili Wang[2,3], Xucun Ma[2,3,*], Qi-Kun Xue[2,3,*], Jimin Zhao[1,*]

* Corresponding author. E-mail: xucunma@mail.tsinghua.edu.cn; qkxue@mail.tsinghua.edu.cn; jmzhao@iphy.ac.cn

### S1. Optimized number of the FeTe capping layer

We have performed control experiment to find the optimized number of the capping layers. In Fig. S1a we show the transient differential reflectivity of the 1UC FeSe covered with different numbers of FeTe protective layers. When the number of protective layer increases from 1 to 2, the transient trace of differential reflection almost remains the same. However, when the number increases to 4, the dynamics becomes very different—the slow component becomes more prominent. When the number increases to 10, the time-resolved signal becomes completely different and the negative component almost disappears. In Fig. S1b, we compare the transient signal of the 1UC FeSe covered by 10UC FeTe sample with the 10UC FeTe only sample. The two signals show almost no difference, thus the FeSe signal is completely concealed by the 10UC FeTe. The thinner the FeTe layer is, the more prominent the signal coming from the FeSe can be detected. To compound with preservation from ambient damage, we chose 2UC FeTe as the protective layer. This control experiment demonstrates that the signals shown in Fig. 1-3 do come from the FeSe layer, thus demonstrating that our ultrafast spectroscopy is single-layer and interface sensitive, being able to detect phase transitions in layered quantum materials. Experiments on a sample of 1UC FeTe on 1UC FeSe have also been done, yielding similarly good experimental data.

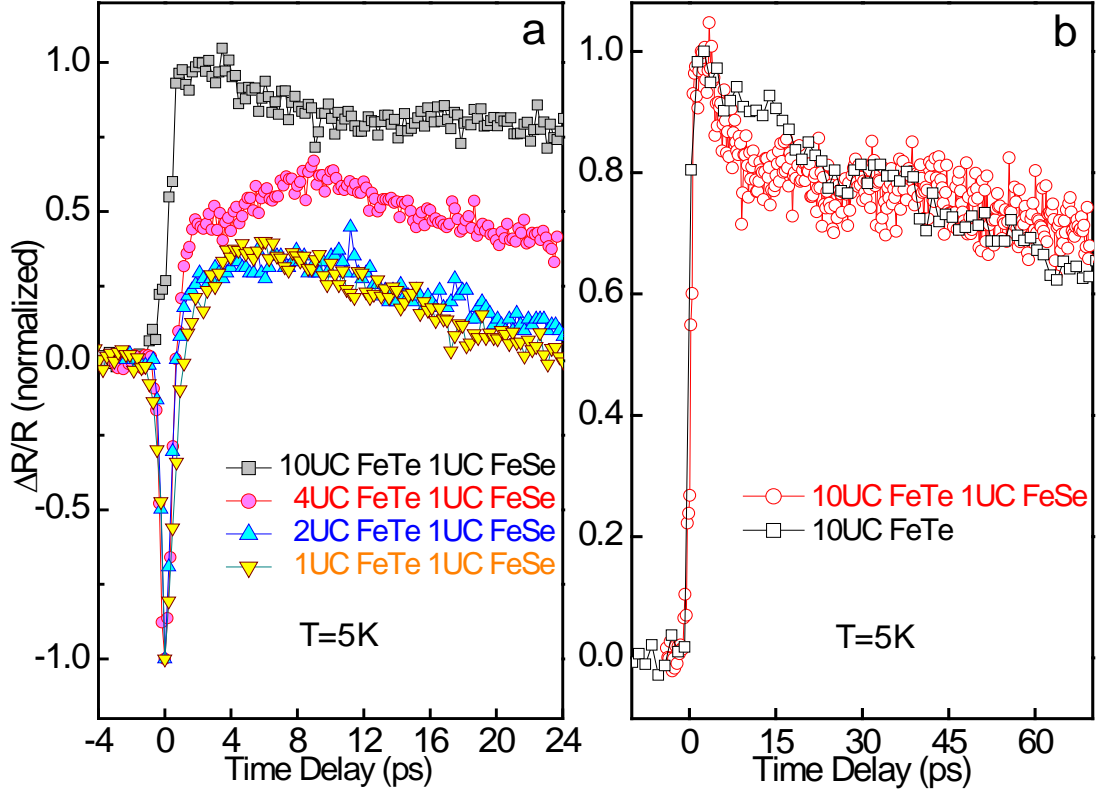

**Fig. S1. Effect of the number of capping layers. a,** The transient photo-induced differential reflectivity of 1UC FeSe covered with different number of FeTe protective layers (at 4.5 K). **b,** Control experiment on samples of 10UC FeTe with and without the 1UC FeSe (at 4.5 K). When the 1UC FeSe sample is covered with 10UC FeTe film, the quasiparticle dynamics of the 1UC FeSe sample cannot be detected.

### S2. Measuring the $T_e$ value from the excitation fluence

The electron temperature $T_e$ is derived from the laser fluence. After the pulse excitation, we have for the single layer sample $T_e = \sqrt{T_i^2 + 2(1-R)F(1-\alpha_{\text{FeTe}})^2 \alpha_{\text{FeSe}}/\gamma d}$, where $T_i$ is the initial sample temperature before excitation, $R$ is the reflectivity, $F$ is the pump fluence, $\alpha$ is the absorbance, $\gamma$ is the linear coefficient of the electron contribution to the total specific heat per unit volume, and $d$ is the effective thickness of the FeSe layer. This can be seen as a generalization for the bulk material case [see Ref. 1 of SI]. We calculate that $T_e$ is 902 K, where we have measured the reflectivity $R = 0.17$, estimated that the absorbance of 1UC FeSe is $\alpha_{FeSe} = 0.0227$ (using the thickness $d = 0.55$ nm and bulk skin depth $l_s = 24$ nm [15]), used $\gamma = 5.73$ mJmol$^{-1}$K$^{-2}$ [see Ref. 2 of SI], taken $F = 268$ $\mu$J/cm$^2$, and assumed $\alpha_{FeTe} \approx \alpha_{FeSe}$.

**S3. Acoustic phonon in 2UC FeTe on STO sample**

Similar to the observation of the acoustic phonon mode in the 10 UC FeTe on STO sample, we have also observed identical acoustic phonon mode in another 2UC FeTe on STO sample using the same experimental setup. The result is illustrated in Fig. S3. In the upper panel, ultrafast dynamics at three different temperatures below and above the $T_c$ are displayed, respectively. In the lower panel excluding the single-exponential decay, the coherent oscillations are clearly shown. In the inset we show the Fourier transformation of the time-domain data, which yields an average phonon frequency of ~0.049 THz (0.203 meV). This is the same as that of the 10UC

FeTe sample. The corresponding period of the acoustic phonon wave is 20.4 ps. The lifetimes of this acoustic phonon are 274 ps at 4.5 K, 262 ps at 24 K, and 200 ps at 110 K, respectively.

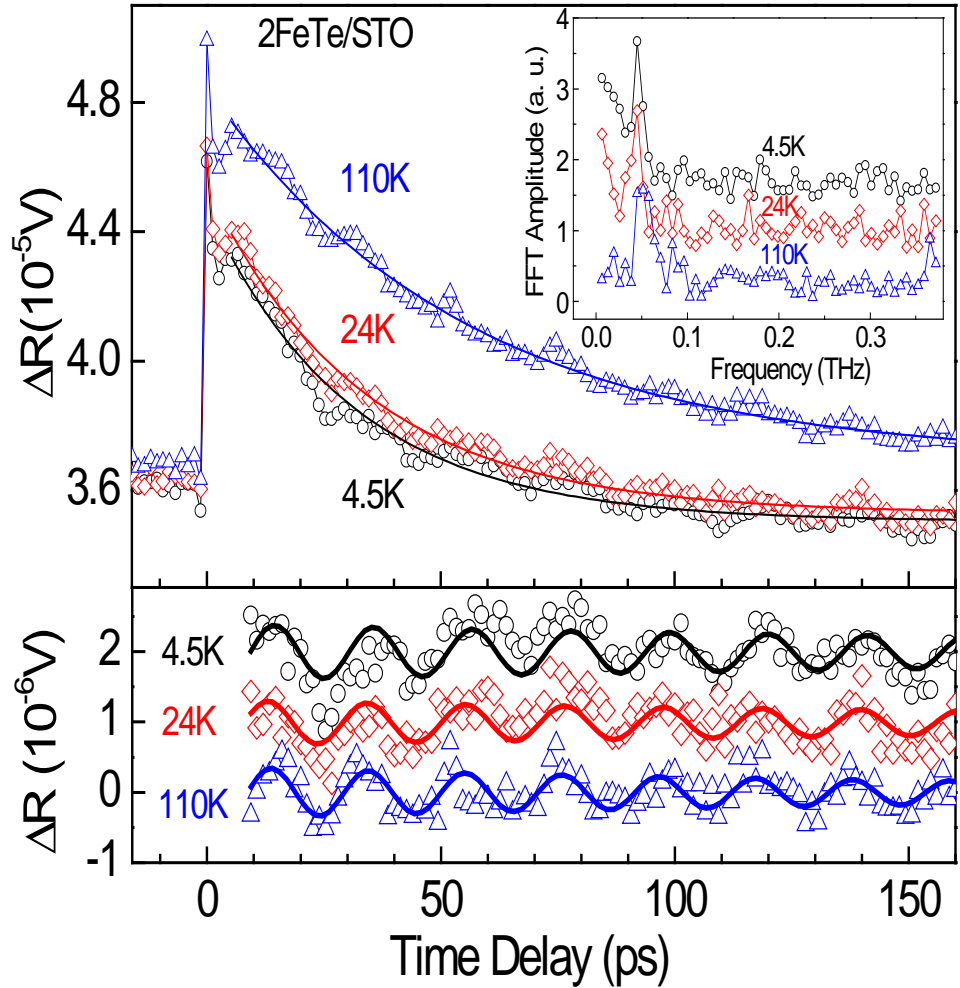

**Fig. S3. Acoustic Phonon in 2UC FeTe on STO**. Upper panel: The transient differential reflections for the 2UC FeTe on $SrTiO_3$ sample at 4.5 K (black), 24 K (red), and 110 K (blue), respectively. The solid lines are single-exponential decay fits. Lower panel: the coherent acoustic phonon oscillation derived by subtracting the exponential decay. Inset: the FFT of acoustic phonon oscillations in the lower panel.

**S4. Pump power dependence of $|\triangle R/R|_{max}$**

We investigated the pump power dependence of the ultrafast dynamics at several different temperatures below $T_c$. The results are summarized in Fig. S4. The y-axis is the maximum absolute value of the differential reflection. The pump fluences are 0.134 mJ/cm$^2$, 0.187 mJ/cm$^2$, 0.267 mJ/cm$^2$, 0.343 mJ/cm$^2$, 0.446 mJ/cm$^2$, 0.58 mJ/cm$^2$, 0.8 mJ/cm$^2$, and 1.1 mJ/cm$^2$, respectively. Note that the slopes of all the dash lines are equal to 1 and there is no interception in the linear plot. Thus the $|\triangle R/R|_{max}$ is proportional to the excitation fluence $\Phi$. Therefore, the density of photo-excited quasi-particles is proportional to laser fluence, $n \propto \Phi$. At all temperatures this relation holds. This manifests that the 1UC FeSe superconductor is node-less, which is consistent with the observation in angle-resolved photoemission spectroscopy.

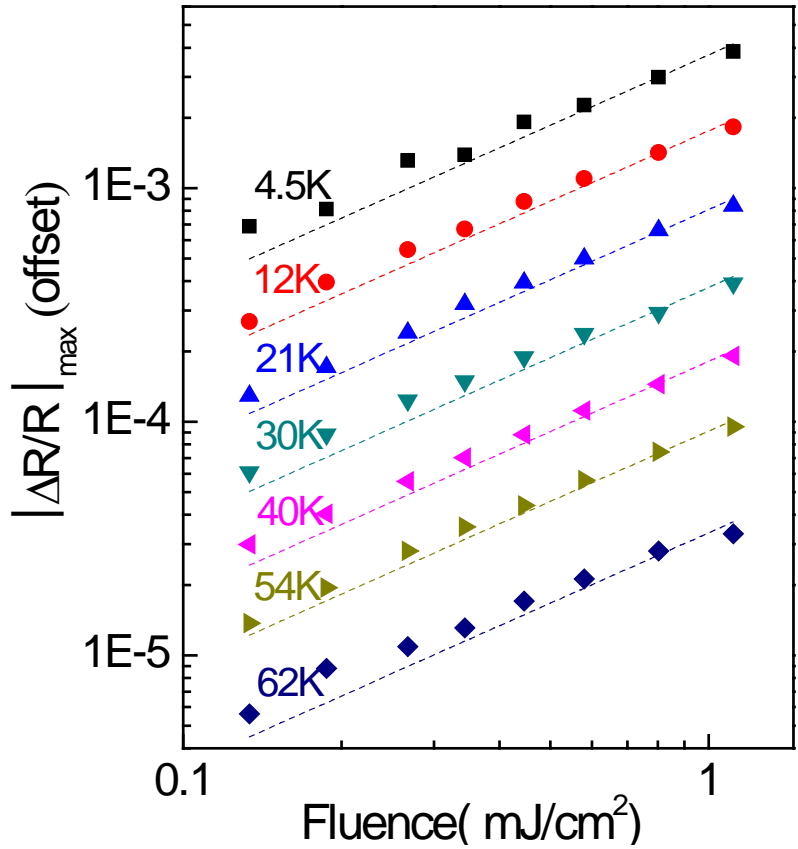

**Fig. S4. The maximum absolute value of differential reflection $|\triangle R/R|_{max}$ as a**

**function of pump fluence**. The data at different temperatures are offset for clarity, respectively. The dash lines are slop = 1 fits to the data.

## S5. Antiferromagnetic to paramagnetic phase transition in FeTe layer

The peak of $\tau_{slow}$ around 103 K in Fig. 3d (light blue color stripe) has no corresponding amplitude decrease in Fig. 3c. Hence it is unlikely a SC transition. The 160 ps lifetime is nearly identical to that for the 10UC FeTe/STO sample (Fig. 4a). Hence we ascribe this peak to the carrier dynamics in the FeTe capping layer, which may correspond to the antiferromagnetic to paramagnetic phase transition. Even with such a strong background of FeTe, we have observed clear features of FeSe (see the negative component in Fig. S1). This manifests that our method is single-layer sensitive and might also apply to other interface investigations.